\documentclass[conference]{IEEEtran}
\usepackage{cite}
\IEEEoverridecommandlockouts
\usepackage{amsmath,amssymb,amsfonts}
\usepackage{booktabs}
\usepackage{multirow}
\usepackage{algorithmic}
\usepackage{graphicx}
\usepackage{textcomp}
\usepackage{xcolor}
\def\BibTeX{{\rm B\kern-.05em{\sc i\kern-.025em b}\kern-.08em
    T\kern-.1667em\lower.7ex\hbox{E}\kern-.125emX}}
\begin{document}

\title{NEST-RQ: Next Token Prediction \\for Speech Self-Supervised Pre-Training}

\author{\IEEEauthorblockN{Minglun Han,
Ye Bai, Chen Shen, Youjia Huang, Mingkun Huang,\\ Zehua Lin, Linhao Dong, Lu Lu, Yuxuan Wang}
\IEEEauthorblockA{\textit{Seed Team, ByteDance}}}

\maketitle
\renewcommand{\baselinestretch}{0.99} \normalsize

\begin{abstract}
Speech self-supervised pre-training can effectively improve the performance of downstream tasks. However, previous self-supervised learning (SSL) methods for speech, such as HuBERT and BEST-RQ, focus on utilizing non-causal encoders with bidirectional context, and lack sufficient support for downstream streaming models. To address this issue, we introduce the next token prediction based speech pre-training method with random-projection quantizer (NEST-RQ). NEST-RQ employs causal encoders with only left context and uses next token prediction (NTP) as the training task. On the large-scale dataset, compared to BEST-RQ, the proposed NEST-RQ achieves comparable performance on non-streaming automatic speech recognition (ASR) and better performance on streaming ASR. We also conduct analytical experiments in terms of the future context size of streaming ASR, the codebook quality of SSL and the model size of the encoder. In summary, the paper demonstrates the feasibility of the NTP in speech SSL and provides empirical evidence and insights for speech SSL research.
\end{abstract}


\section{Introduction}

In recent years, there has been significant progress in the field of speech self-supervised learning (SSL) technology, which attracts widespread attention from both the academic and industrial communities~\cite{oord2018representation,chung2019unsupervised,baevski2020wav2vec2,chung2021w2v,hsu2021hubert,baevski2022data2vec,chen2022wavlm}. By mining the information from a large amount of unlabeled speech data, speech SSL can provide powerful representations or representation models for downstream speech tasks~\cite{mohamed2022self}. SSL is pushing the performance of downstream tasks to new heights. For instance, speech SSL has facilitated many ASR models to achieve state-of-the-art (SOTA) performance on various benchmarks~\cite{zhang2023google,bai2024seed}. These studies demonstrate the great potential of speech SSL in speech applications.

However, most popular SSL methods, such as Wav2vec 2.0~\cite{baevski2020wav2vec2}, HuBERT~\cite{hsu2021hubert}, Data2vec~\cite{baevski2022data2vec}, focus on exploring non-causal encoders that attend to bidirectional context, and ignore the downstream streaming tasks. Applying the non-causal encoder to streaming models not only requires modifying the encoder~\cite{li2023dynamic,fu2023ufo2} and the training strategy~\cite{cao2021improving,doutre2021improving,yu2021dual,yao2021wenet}, but may also result in sub-optimal performance. Therefore, it is not easy to directly adapt the popular SSL methods to downstream streaming tasks. Furthermore, there are a limited number of studies on self-supervised learning (SSL) specifically for streaming models in downstream tasks~\cite{chiu2022self,fu2023ufo2}. Although BEST-RQ~\cite{chiu2022self} explores speech self-supervised pre-training for streaming ASR models, some of its conclusions are still unclear and needed to be further investigated due to the adoption of different settings for streaming and non-streaming ASR tasks. Among SSL methods that can adapt to streaming tasks, CPC~\cite{oord2018representation} and APC~\cite{chung2019unsupervised} use the causal encoder as the backbone, and adopt contrastive predictive coding and auto-regressive predictive coding as pre-training objectives, respectively. However, these methods have not been evaluated on current popular end-to-end ASR models, and there is a gap in both structure and performance compared to the current popular SSL methods.

In this paper, we propose the \textit{\textbf{NE}xt token prediction based \textbf{S}peech pre-\textbf{T}raining with \textbf{R}andom-projection \textbf{Q}uantizer} (\textbf{NEST-RQ}), a novel speech SSL method that uses next token prediction (NTP) as the pre-training objective. Currently, the large language model (LLM) \cite{zhao2023survey,radford2019language,Brown2020LanguageMA} that relies on NTP-based SSL has achieved great success, and has been widely explored and applied in many fields~\cite{bai2024seed,liu2024visual,chen2023x,tang2023salmonn,chu2023qwen}. However, due to the continuous nature of speech, it is not easy to apply NTP to speech SSL. With the help of the random-projection quantizer (RQ)~\cite{chiu2022self}, we can convert the speech into the token sequence. Thus, we can introduce the NTP task in speech SSL and further optimize it based on the characteristics of speech. This paper verifies the feasibility of NTP in speech SSL, and brings more inspiration for joint auto-regressive modeling of audio and text. Our contributions are listed as follows: 
\begin{enumerate}
    \item We propose a novel speech SSL method named NEST-RQ. First, the encoder in NEST-RQ is set to the causal structure. Then, RQ converts the continuous speech features into a sequence of discrete tokens. Finally, the causal encoder takes continuous speech features as inputs and uses the output of the encoder at each frame to predict the tokens of multiple subsequent frames;
    \item We show the effectiveness of NEST-RQ on a large-scale dataset that covers 300,000 hours of unlabeled speech and 30,000 hours of labeled speech. NEST-RQ achieves comparable performance to BEST-RQ on non-streaming ASR task and outperforms BEST-RQ on streaming ASR task. NEST-RQ inherits the simplicity of BEST-RQ while maintaining performance;
    \item We conduct analytical experiments to explore the performance of NEST-RQ in terms of the codebook quality in SSL, encoder size in SSL, and future context size in streaming ASR. Experiments show that NEST-RQ brings consistent improvements on streaming ASR across different settings.
\end{enumerate}

\section{Methodology}
\subsection{Preliminaries}
\subsubsection{SSL Method}
BEST-RQ~\cite{chiu2022self} is an effective approach for speech SSL. Specifically, the method masks some segments of speech features and input the speech features into the speech encoder. The encoder learns to predict the masked segments based on the unmasked speech features, and the learning targets are generated by a random-projection quantizer (RQ). The random-projection quantizer projects the speech feature with a randomly-initialized matrix, finds the nearest vector in a randomly-initialized codebook, and uses the index of that vector as the target token. During training, both the projection matrix and the codebook is fixed. Benefiting from the simple quantizer design and the widely-recognized masked prediction task, BEST-RQ shows unique advantages among SSL methods. However, BEST-RQ relies on bidirectional context to predict the tokens of the masked segments, which makes it not easy to adapt to downstream streaming models.

\subsubsection{Downstream ASR Model}
\begin{figure}[t]
  \centering
  \includegraphics[width=0.45\textwidth]{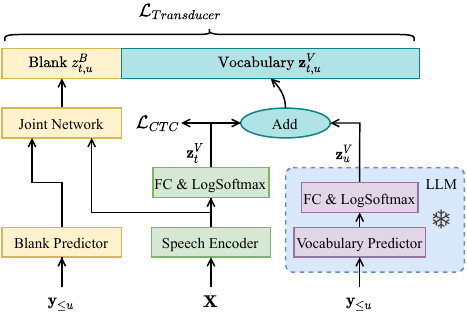}
  \vspace{-10pt}
  \caption{Factorized neural transducer (FNT) with LLM.}
  \label{fig:fnt}
  \vspace{-10pt}
\end{figure}

Factorized neural transducer (FNT)~\cite{chen2022factorized,zhao2023fast} is used in downstream ASR task in this work. By controlling the causality of the encoder, FNT can perform streaming and non-streaming ASR. As shown in Fig.\ref{fig:fnt}, FNT consists of a speech encoder, the joint network, and the vocabulary predictor and the blank predictor. FNT predicts the blank token and vocabulary tokens separately, so that the vocabulary predictor can function as a language model (LM). Thus, inspired by studies~\cite{chen2023x,wu2023decoder,fathullah2024prompting,bai2024seed} that apply LLMs to end-to-end ASR models, we directly use a pre-trained LLM to initialize the vocabulary predictor and its fully-connected layer (FC) for projection, and freeze both of them during training. The output sequence of the encoder is downsampled by a factor of 2 using a convolutional layer, and then used to generate logits for connectionist temporal classification (CTC)~\cite{graves2006connectionist}. Since the LLM keeps frozen, the loss $\mathcal{L}_{LM}$ for LM in original FNT is removed. Thus, the total loss for FNT is the sum of loss $\mathcal{L}_{Transducer}$ for transducer and loss $\mathcal{L}_{CTC}$ for CTC, and can be written as:
$$ \mathcal{L}_{FNT} = \mathcal{L}_{Transducer}+ \mathcal{\lambda}_{CTC}  \ \mathcal{L}_{CTC}.$$

\subsection{NEST-RQ}

\begin{figure}[t]
  \centering
  \includegraphics[width=0.46\textwidth]{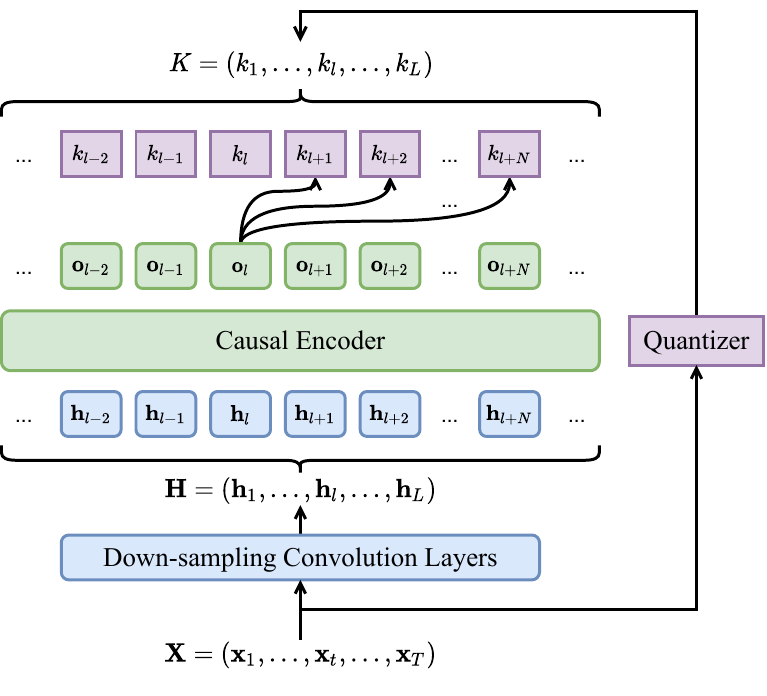}
  \vspace{-10pt}
  \caption{NEST-RQ.}
  \label{fig:nestrq}
  \vspace{-10pt}
\end{figure}

We propose a novel speech SSL method called NEST-RQ. NEST-RQ uses the causal encoder that can only attend to the current frame and the past frames of speech features, and applies NTP as the pre-training task. Both causal encoders and NTP task make NEST-RQ friendlier to downstream streaming models. The training task of NEST-RQ is illustrated in Fig.\ref{fig:nestrq}. After the feature extraction, raw speech can be transformed into speech features $\mathbf{X} = (\mathbf{x}_1,\mathbf{x}_2,...,\mathbf{x}_t, ..., \mathbf{x}_T)$. The convolution layers provides 4 times temporal-dimension reduction for $\mathbf{X}$. After down-sampling, the feature sequence can be denoted as $\mathbf{H}=(\mathbf{h}_{1}, ..., \mathbf{h}_{l}, ...,\mathbf{h}_{L})$. Meanwhile, we use the same RQ in BEST-RQ to generate token sequence $K=(k_1, k_2, .., k_l, ...,k_L)$ from $\mathbf{X}$. Since the down-sampling convolution layers provides 4 times temporal-dimension reduction, RQ combines every 4 frames for projection. After processing input feature sequence $\mathbf{H}$, the conformer module~\cite{gulati2020conformer} outputs state sequence $\mathbf{O}=(\mathbf{o}_{1},\mathbf{o}_{2} ..., \mathbf{o}_{l}, ...,\mathbf{o}_{L})$. Then, we can use $l$-th state $\mathbf{o}_l$ to predict multiple subsequent tokens. Specifically, assuming the current position is $l$, the output state $\mathbf{o}_l$ is used to predict the tokens $\{k_{l+1}, ..., k_{l+n}, ..., k_{l+N}\}$ of the next $N$ tokens with $N$ different prediction heads. The loss at the $l$-th position can be written as 
$-\sum^{N}_{n=1}log\:p \left( k_{l+n} |\mathbf{h}_{\le l} \right)$. Finally, the NEST-RQ loss for one sample can be represented as
$$\mathcal{L}_{{NEST-RQ}} = -\sum^{L}_{l=1}\ \sum^{N}_{n=1}log\ p \left( k_{l+n} |\mathbf{h}_{\le l} \right).$$


\subsection{Encoder Adaptation from SSL to ASR}

\begin{table*}[t]
\centering
\caption{The main experimental results on the large-scale dataset. The column with header ``Average" shows the average CER (\%).}
\resizebox{\linewidth}{!}{%
\begin{tabular}{@{}ll|cccccc|cccccc@{}}
\toprule[2pt]
\multicolumn{2}{l|}{ASR Mode}            & \multicolumn{6}{c|}{Streaming}             & \multicolumn{6}{c}{Non-Streaming}          \\ \midrule
\multicolumn{1}{l|}{SSL Method} & SSL Causality & test-1 & test-2 & test-3 & test-4 & \multicolumn{1}{c|}{test-5} & Average & test-1 & test-2 & test-3 & test-4 & \multicolumn{1}{c|}{test-5} & Average \\ \midrule
\multicolumn{1}{l|}{BEST-RQ} & {\texttt{NC-A}},\ {\texttt{NC-C}} & 13.3 & 13.9 & 12.3 & 16.7 & \multicolumn{1}{c|}{12.3} & 13.7 & 10.0 & 10.4 & 9.2  & 12.5 & \multicolumn{1}{c|}{10.6} & 10.5 \\
\multicolumn{1}{l|}{BEST-RQ} & {\texttt{C-A}},\ {\texttt{NC-C}} & 13.5 & 13.9 & 12.3 & 16.7 & \multicolumn{1}{c|}{12.4} & 13.8 & 10.2 & 10.5 & 9.3  & 12.7 & \multicolumn{1}{c|}{10.7} & 10.7 \\
\multicolumn{1}{l|}{BEST-RQ} & {\texttt{C-A}},\ {\texttt{C-C}}   & 14.3 & 14.7 & 13.2 & 17.8 & \multicolumn{1}{c|}{12.8} & 14.5 & 10.5 & 11.2 & 10.1 & 13.6 & \multicolumn{1}{c|}{10.9} & 11.3 \\ \midrule
\multicolumn{1}{l|}{NEST-RQ} & {\texttt{C-A}},\ {\texttt{C-C}} & \textbf{12.7} & \textbf{13.2} & \textbf{11.4} & \textbf{15.6} & \multicolumn{1}{c|}{\textbf{12.0}} & \textbf{13.0} & \textbf{10.0} & \textbf{10.2} & \textbf{9.0}  & \textbf{12.3} & \multicolumn{1}{c|}{\textbf{10.5}} & \textbf{10.4} \\ \bottomrule[2pt]
\end{tabular}%
}
\vspace{-13pt}
\label{tab:main_exp_inhouse}
\end{table*}

Non-causal encoders are usually used as the backbone in past SSL studies~\cite{baevski2020wav2vec2,hsu2021hubert,chiu2022self}. ``Non-causal'' means that the encoder uses both past and future context, while ``causal'' means that the encoder can only use past context. For ASR task, non-causal encoders are suitable for non-streaming models, while causal encoders are suitable for streaming models. In this section, we focus on adapting the pre-trained causal or non-causal encoder for streaming and non-streaming ASR. 

As the basic component of the encoder in this work, the conformer block~\cite{gulati2020conformer} includes two modules that affect the causality of the encoder: attention module and depth-wise convolution module. Here, we denote the non-causal attention module that attends to the current frame, all past and all future frames, as {\texttt{NC-A}}, and denote the causal attention module that only attend to the current and the past frames, as {\texttt{C-A}}. Similarly, the non-causal convolution with a kernel that covers the past, the current and the future frames is denoted as {\texttt{NC-C}}, while the causal convolution with a kernel that only covers the current and the past frames is denoted as {\texttt{C-C}}.

By controlling the attention mask, we can easily switch between {\texttt{NC-A}} and {\texttt{C-A}}. However, adjusting the causality of the convolution is more complicated. Here, we should consider two situations: 1) When transforming {\texttt{NC-C}} of the pre-trained encoder into {\texttt{C-C}} of the ASR encoder, we remove the right-side parameters of the convolution kernel. For example, if the original {\texttt{NC-C}} has a kernel with size $\left( 2m+1 \right) $, the kernel of its corresponding {\texttt{C-C}} will have size $\left ( m+1 \right) $. 2) When transforming {\texttt{C-C}} of the pre-trained encoder into {\texttt{NC-C}} of the ASR encoder, the convolution kernel with size $\left ( m+1 \right) $ will be expanded at the right side with parameters initialized with Xavier uniform distribution, and finally has size $\left( 2m+1 \right) $.

\section{Experimental Settings}
\label{sec:exp_settings}

\subsection{Data}
The SSL of the encoder uses 300,000 hours of in-house unlabeled speech data. The supervised fine-tuning (SFT) of the ASR model uses 30,000 hours of in-house ASR data, and the ASR test sets covers five subsets. All data cover multiple challenging scenarios, including video, live, etc. All input speech features are 80-dimensional log-mel filterbank coefficients, and each frame has the stride of 10ms. The metric used for ASR evaluation is character error rate (CER).

\subsection{Model}

We use encoders with different sizes: 0.1B, 0.3B, 0.6B. Both 0.1B and 0.6B encoders have almost the same structure with that in ~\cite{zhang2020pushing}. However, the non-causal convolution kernel size of 0.1B is set to 31 $\left( 2 \times m + 1 , m=15 \right)$ in this work. 0.3B encoder has the same settings with 0.6B encoder but has half number of conformer blocks. Among three encoders, 0.3B is the default choice. For non-streaming ASR, the attention module is {\texttt{NC-A}}, and the convolution is {\texttt{NC-C}}. For streaming ASR, the attention in the bottom $M$ blocks attend to the current frame, the past frames and 1 future frame, while the attention in other blocks keep strict causality as {\texttt{C-A}}. The convolution is {\texttt{C-C}}. $M$ controls the future context size in streaming ASR, and $M$ is set to 3 by default. The LLM used in FNT is an in-house LLM with over 1 billion parameters.

\subsection{Training and Inference} 

\subsubsection{SSL}

We conduct training for about 2 epochs with the batch size of 1.5 hours. We use the transformer learning rate (LR) scheduler with warm-up of 8k steps and peak LR 3e-4. For NEST-RQ, we set the $N$ in multi-token prediction to 5. For BEST-RQ, the mask length and mask ratio $p$ are set to 400ms and 0.012, respectively. The vocabulary size of the codebook in RQ is set to 1024 and the dimension of vectors in codebook is set to 16. 

\subsubsection{SFT}

The pre-trained encoder in SSL stage is used to initialize the speech encoder in downstream ASR model. During training, we freeze the LLM and update other modules in ASR model. $\lambda_{CTC}$ is set to 1.0. We train the ASR model for 3 epochs with the batch size of 1.5 hours. We use the linear LR scheduler, with the LR increasing to 1e-4 until 6k steps and then gradually decaying. The frequency masking and time masking in SpecAugment~\cite{park2019specaugment} are also applied. During inference, we conduct beam search with beam size 10.

\section{Experimental Results}
\label{sec:typestyle}

\subsection{Results on the Large-Scale Dataset}

\begin{table}[t]
\centering
\caption{The effects of the number $N$ of tokens in multi-token prediction of NEST-RQ on performance. The experiments are conducted on 0.1B encoder, and the metric is average 
 CER (\%). }
\resizebox{\linewidth}{!}{%
\begin{tabular}{l|ccccccc}
\toprule[2pt]
$N$ & 1    & 3    & 5    & 7    & 10   & 20   & 50  \\ \midrule
Streaming              & 15.9 & 14.9 & \textbf{14.7} & 14.8 & 15.1 & 15.9 & 16.6  \\
Non-Streaming          & 12.0 & 11.5 & \textbf{11.4} & 11.5 & 11.7 & 12.2 & 12.6   \\ \bottomrule[2pt]
\end{tabular}%
}
\vspace{-13pt}
\label{table:number_tokens}
\end{table}

\begin{table*}[ht]
\centering
\caption{The analytical results with different codebooks used in speech SSL. Both Q1 and Q2 are quantizers with better codebooks. The column with header ``Average" shows the average CER (\%).}
\resizebox{0.9\linewidth}{!}{%
\begin{tabular}{l|cccccc|cccccc}
\toprule[2pt]
ASR Mode    & \multicolumn{6}{c|}{Streaming}                                  & \multicolumn{6}{c}{Non-Streaming}                              \\ \midrule
SSL Method & test-1 & test-2 & test-3 & test-4 & \multicolumn{1}{c|}{test-5} & Average & test-1 & test-2 & test-3 & test-4 & \multicolumn{1}{c|}{test-5} & Average \\ \midrule
BEST-RQ     & 13.3 & 13.9 & 12.3 & 16.7 & \multicolumn{1}{c|}{12.3} & 13.7 & 10.0 & 10.4 & 9.2 & 12.5 & \multicolumn{1}{c|}{10.6} & 10.5 \\
NEST-RQ     & 12.7 & 13.2 & 11.4 & 15.6 & \multicolumn{1}{c|}{12.0} & 13.0 & 10.0 & 10.2 & 9.0 & 12.3 & \multicolumn{1}{c|}{10.5} & 10.4 \\ \midrule
BEST-Q1  & 12.3 & 12.6 & 10.7 & 15.0 & \multicolumn{1}{c|}{11.7} & 12.5 & \textbf{8.8}  & \textbf{9.1}  & 7.6 & 11.0 & \multicolumn{1}{c|}{10.0} & 9.3  \\
NEST-Q1  & 11.6 & 12.3 & 10.0 & 14.5 & \multicolumn{1}{c|}{11.4} & 12.0 & 9.2  & 9.6  & 7.9 & 11.3 & \multicolumn{1}{c|}{10.1} & 9.6  \\ \midrule
BEST-Q2 & 12.0 & 12.2 & 10.3 & 14.7 & \multicolumn{1}{c|}{11.5} & 12.2 & 8.8  & 9.2  & \textbf{7.5} & \textbf{10.9} & \multicolumn{1}{c|}{\textbf{9.8}}  & \textbf{9.2}  \\
NEST-Q2 & \textbf{11.5} & \textbf{12.1} & \textbf{9.9}  & \textbf{14.2} & \multicolumn{1}{c|}{\textbf{11.3}} & \textbf{11.8} & 9.2  & 9.5  & 7.9 & 11.2 & \multicolumn{1}{c|}{10.1} & 9.6  \\ \bottomrule[2pt]
\end{tabular}%
}
\vspace{-13pt}
\label{tab:codebook}
\end{table*}

As shown in Table \ref{tab:main_exp_inhouse}, the main experiments on the large-scale dataset are conducted on the 0.3B encoder. On non-streaming ASR, BEST-RQ and NEST-RQ show comparable performance, while on streaming ASR, NEST-RQ outperforms BEST-RQ. Besides, compared to the pre-training methods (BEST-RQ with {\texttt{C-A}},\ {\texttt{NC-C}}, and BEST-RQ with {\texttt{C-A}}, \ {\texttt{C-C}}) for streaming models in BEST-RQ~\cite{chiu2022self}, NEST-RQ still demonstrates further improvements. NEST-RQ has several advantages over BEST-RQ: 1) During training, NEST-RQ is more sample-efficient than BEST-RQ, as BEST-RQ only predicts on the masked segments; 2) NEST-RQ can better adapt to streaming tasks due to the use of the causal encoder. Last but not least, NEST-RQ inherits the simplicity of BEST-RQ. If not specified, BEST-RQ is equipped with {\texttt{NC-A}} and {\texttt{NC-C}} in all subsequent experiments.

In addition, we investigate the effect of the number of predicted tokens in NEST-RQ in Table \ref{table:number_tokens}. We use the values from $\{1,3,5,7,10,20,50\}$ to explore the optimal $N$ in multi-token prediction. When $N$ is set to 5, the model achieves the best performance. Due to the smoothness of speech signals, the adjacent speech frames are similar in short period. Thus, when $N$ is too small ($N=1$), the SSL training task tends to be trivial, making it difficult for the model to learn meaningful and high-quality representations. We set $N$ to 5 in all other NEST-RQ experiments.

\vspace{-5pt}

\subsection{Analytical Results in Different Settings}
\subsubsection{Analysis with Different Codebooks}

\begin{table}[t]
\centering
\caption{The analytical results with different encoder sizes. The metric is average CER (\%).}
\resizebox{0.90\linewidth}{!}{%
\begin{tabular}{l|ccc|ccc}
\toprule[2pt]
ASR Mode           & \multicolumn{3}{c|}{Streaming} & \multicolumn{3}{c}{Non-Streaming} \\ \midrule
Size (B) & 0.1     & 0.3     & 0.6     & 0.1       & 0.3      & 0.6     \\ \midrule
BEST-RQ               & 15.8    & 13.7    & \textbf{12.9}  & 11.8      & 10.5     & \textbf{9.7}     \\
NEST-RQ               & 14.7    & 13.0    & \textbf{12.1}    & 11.4      & 10.4     & \textbf{9.6}     \\ \bottomrule[2pt]
\end{tabular}%
}
\vspace{-13pt}
\label{tab:scalability-model_sizes}
\end{table}

We conduct analytical experiments with different codebooks. The random-projection quantizer is still represented as RQ. In addition, we introduce two extra quantizers Q1 and Q2 with high-quality codebooks. Specifically, the codebook in Q1 is generated by clustering the outputs of a non-causal encoder, which is trained with BEST-RQ on 3.5 million hours of unlabeled speech, into 1024 centroids. The codebook in Q2 is generated by clustering the outputs of a non-causal encoder, which is trained with BEST-RQ on 7.5 million hours of unlabeled speech, into 32768 centroids. As shown in Table \ref{tab:codebook}, with the help of more powerful codebooks, both BEST and NEST can achieve further improvements, verifying the scalability of the NEST method with different codebooks. However, we observe that NEST is worse than BEST on non-streaming ASR task. We conjecture that this is because the high-quality codebooks are derived from the non-causal encoder trained with BEST-RQ, which is more compatible with BEST.

\subsubsection{Analysis with Different Encoder Sizes}
The experiments are extended to different encoder sizes. We explore 3 different model sizes mentioned in section~\ref{sec:exp_settings}. From Table \ref{tab:scalability-model_sizes}, we observe that: 1) The performance of BEST-RQ and NEST-RQ on both streaming and non-streaming models gradually improves as the encoder size increases; 2) Compared to BEST-RQ, NEST-RQ consistently brings relative error rate reduction of 5\% to 7\% on streaming ASR and maintains comparable non-streaming performance across different encoder sizes.

\subsubsection{Analysis with Different Future Context Sizes}

In streaming ASR, the attention modules of the bottom 3 conformer blocks attend to the past frames, the current frame and the next frame by default. The more blocks that can attend to the next frame, the larger the receptive field for future context. Therefore, by controlling the number of conformer blocks $M$ that attend to the next frame, we can control the future context size of the encoder. Table \ref{tab:future_context_size} shows the streaming ASR performance when $M$ takes values from $\{0,1,3,5,7\}$. In all settings, NEST-RQ shows better performance. As the future context size increases, the model tends to be more non-streaming, resulting in increased recognition latency and decreased performance gain.

\begin{table}[t]
\centering
\caption{The analytical results with different future context sizes. The metric is average CER (\%).}
\resizebox{0.78\linewidth}{!}{%
\begin{tabular}{l|ccccc}
\toprule[2pt]
$M$ & 0    & 1    & 3    & 5    & 7    \\ \midrule
BEST-RQ & 14.7 & 14.2 & 13.7 & 13.4 & \textbf{13.2} \\
NEST-RQ & 13.7 & 13.5 & 13.0 & 12.8 & \textbf{12.7} \\ \bottomrule[2pt]
\end{tabular}%
}
\vspace{-13pt}
\label{tab:future_context_size}
\end{table}

\section{Conclusion}

In this paper, we propose a novel speech SSL method named NEST-RQ. NEST-RQ employs a causal encoder and predicts the tokens of the subsequent positions in the sequence as the SSL task. The causal encoder makes NEST-RQ more friendly to downstream streaming tasks. Experiments on the large-scale dataset show that NEST-RQ achieves comparable performance to BEST-RQ on non-streaming ASR task and outperforms BEST-RQ on streaming ASR task. In the future, we will explore the performance of the NEST-RQ on more downstream tasks and transfer it into the joint auto-regressive modeling of speech and text.

\bibliographystyle{IEEEtran}
\bibliography{conference_101719}

\end{document}